\documentclass[9pt,conference]{IEEEtran}
\usepackage{cite}

\ifCLASSINFOpdf
\else
\fi
\begin{document}
%
\title{Hyperspectral fluorescence microscopy based \\on Compressive Sampling}

\author{\IEEEauthorblockN{Makhlad Chahid\IEEEauthorrefmark{1}\IEEEauthorrefmark{2}, J\'{e}rome Bobin\IEEEauthorrefmark{3}, Hamed Shams Mousavi\IEEEauthorrefmark{1}\IEEEauthorrefmark{2}, Emmanuel Candes\IEEEauthorrefmark{4}, Maxime Dahan\IEEEauthorrefmark{5} and Vincent Studer\IEEEauthorrefmark{1}\IEEEauthorrefmark{2}}
\IEEEauthorblockA{\IEEEauthorrefmark{1}Universit\'{e} Bordeaux 2, Interdisciplinary Institute for Neuroscience, UMR 5297, F-33000 Bordeaux, France\\}
\IEEEauthorblockA{\IEEEauthorrefmark{2}CNRS, Interdisciplinary Institute for Neuroscience, UMR 5297, F-33000 Bordeaux, France\\}
\IEEEauthorblockA{\IEEEauthorrefmark{3}CEA Saclay, IRFU/SEDI-SAP, F-91191 Gif sur Yvette Cedex, France}
\IEEEauthorblockA{\IEEEauthorrefmark{4}Departments of Mathematics, of Statistics, and of Electrical Engineering\\
Standford University, Stanford, California 94305}
\IEEEauthorblockA{\IEEEauthorrefmark{5}Laboratoire Kastler Brossel, CNRS, UMR 8552, \'{E}cole Normale Sup\'{e}rieure\\
Universit\'{e} Pierre et Marie Curie-Paris 6, 75005 Paris, France}}


\maketitle

\begin{abstract}
The mathematical theory of compressed sensing (CS) asserts that one can acquire signals from measurements whose rate is much lower than the total bandwidth. Whereas the CS theory is now well developed, challenges concerning hardware implementations of CS-based acquisition devices---especially in optics---have only started being addressed. This paper presents an implementation of compressive sensing in fluorescence microscopy and its applications to biomedical imaging. Our CS microscope combines a dynamic structured wide-field illumination and a fast and sensitive single-point fluorescence detection to enable reconstructions of images of fluorescent beads, cells, and tissues with undersampling ratios (between the number of pixels and number of measurements) up to 32. We further demonstrate a hyperspectral mode and record images with 128 spectral channels and undersampling ratios up to 64, illustrating the potential benefits of CS acquisition for higher-dimensional signals, which typically exhibits extreme redundancy. Altogether, our results emphasize the interest of CS schemes for acquisition at a significantly reduced rate and point to some remaining challenges for CS fluorescence microscopy.
\end{abstract}

\begin{IEEEkeywords}
biological imaging, compressed sensing, computational imaging, sparse signals.
\end{IEEEkeywords}



%
\IEEEpeerreviewmaketitle

\section{Introduction}
In fluorescence microscopy, one can distinguish two kinds of imaging approaches, wide-field and raster scan microscopy, differing by their excitation and detection scheme\cite{mertz2009introduction}. In both imaging modalities the acquisition is independent of the information content of the image. Rather, the number of acquisitions N, is imposed by the Nyquist-Shannon theorem. However, in practice, many biological images are compressible (or, equivalently here, sparse), meaning that they depend on a number of degrees of freedom K that is smaller that their size N. Recently, the mathematical theory of compressed sensing (CS)\cite{1614066,1580791} has shown how the sensing modality could take advantage of the image sparsity to reconstruct images with no loss of information while largely reducing the number M of acquisition.

\subsection{Compressive Fluorescence Microscopy (CFM)}
Here we present a novel fluorescence microscope designed along the principles of CS.  It uses a DMD to create structured wide-field excitation patterns and a sensitive point-detector to measure the emitted fluorescence. On sparse fluorescent samples (beads and fluorescently labeled living cells), we could achieve compression ratio N/M of up to 64, meaning that an image can be reconstructed with a number of measurements of only 1.5\% of its pixel number.

\subsection{Hyperspectral Imaging via CFM}
Hyperspectral imaging is defined as the combined acquisition of spatial and spectral information. In biological imaging, a growing range of applications such as the study of protein localization and interactions require quantitative approaches that analyze several distinct fluorescent molecules at the same time in the same sample\cite{Zimmermann200387}. These applications are in fact becoming ever more common with the availability of an increasing panel of fluorescent dyes and proteins with emission ranging from the UV to the far red\cite{Giepmans14042006}.
We demonstrate how CS acquisition schemes can be extended to an hyperspectral imaging system. We could acquire fluorescence images, with 128 different spectral channels, with a compression ratio of up to 128.

\section{Conclusion}

We have developed an imaging approach based on the concepts of CS theory which, on samples relevant for biological imaging, allows the reconstruction of fluorescence images with undersampling ratios up to 64. While our results constitute a significant gain over undersampling ratios achieved in prior CS-based imaging approaches, several factors (computational, instrumental, or noise-related) still contribute to limit the current performances of CFM. We finally discuss strategies to further reduce the number of acquisition by taking into account the sample sparsity, not only in the spatial but also in the spectral domain.


\section*{Acknowledgment}

Deepak Nair is gratefully acknowledged for providing the Zyxin-mEOS2 transfected COS7 cells. This work was supported by a grant from the European Aeronautic Defence and Space Foundation. M.C. is supported by the EADS foundation and the Conseil R\'{e}gional d'Aquitaine. E.C. is partially supported by National Science Foundation via grant CCF-0963835 and the 2006 Waterman Award, by Air Force Office of Scientific Research under Grant FA9550-09-1-0643 and by Office of Naval Research under Grant N00014-09-1-0258.




\bibliographystyle{IEEEtran}
\bibliography{IEEEabrv,mybibfileIEEE}
%
%
%

\end{document}